\documentclass[11pt,twoside]{article}

\usepackage{asp2004}
\usepackage{epsf}
\usepackage{graphicx}
\usepackage{psfig}
\usepackage{lscape}

\begin{document}
\title{Binaries as astrophysical laboratories: an overview}   
\author{Carla Maceroni}   
\affil{INAF--Osservatorio Astronomico di Roma, via Frascati 33, I-00040 Monteporzio C. (RM) - Italy}    

\begin{abstract} 

The study of binary stars is worth to undertake not only to learn more about the 
properties  of binaries as such, but also because binaries are
multi-purpose astrophysical tools.  This paper reviews some of the
ways this effective ``tool" can be used, focusing on fundamental parameter
determination, tests of theoretical models, and the recent contribution of binary
stars to establish the  distance to the Magellanic Clouds, and therefore, 
the first rungs of the cosmological distance ladder. 
\end{abstract}


\section{Introduction: binarity as a  benefit}   

A recurrent activity of the astrophysicist involved in binary star 
research is  ``advertising"  binaries as effective
astrophysical tools. 
It could seem hardly necessary, as the  the role
of binaries in laying the foundations of stellar astrophysics is, in principle,
very well known.  Too often, however, the complexity that
a binary configuration can add to observations, to data analysis and interpretation,
overtake the large amount of information that  could, nevertheless, be extracted from
binarity.
 
 To fully exploit the binary assets accurate, and sometimes lengthy, observations are
needed. In some cases (as for spectroscopy of extragalactic binaries) world-class
instrumentation is required, that is rather difficult to obtain for the necessary
number of nights. 
Therefore, it is important to stress that the efforts to better understand  binary
stars   yield  as well ``by-product" useful results  for stellar astrophysics. 

\begin{table*}[!ht]
\label{atglance}
\caption{Parameter determination at glance}
\smallskip
\begin{center}
{\small
\begin{tabular}{lcccccc}
\tableline
\noalign{\smallskip}
Element & \multicolumn{2}{c}{AB}&& \multicolumn{2}{c}{SB}& EB \\
\noalign{\smallskip}
\cline{2-3}
\cline{5-6}
\noalign{\smallskip}
& VB & IB && SB1 & SB2 & \\
\noalign{\smallskip}
\tableline
\noalign{\smallskip}
$a$		&$a \arcsec$  &$a \arcsec$ &&$a_1\sin i$	& $a \sin i$ & N	\\
\noalign{\smallskip}
$e$		& Y & Y && Y & Y	&	Y\\
\noalign{\smallskip}
$P, T_0$& Y	& Y && Y	& Y &	Y\\
\noalign{\smallskip}
$i$		& Y	& Y	&& N	& N	& Y	\\
\noalign{\smallskip}
$\omega$& Y &	Y	&&Y	&Y	& Y  \\
\noalign{\smallskip}
$\Omega$& $\pm 180\deg$ & $\pm 180\deg$  && N	& N	&  N \\
\noalign{\smallskip}
$m_1, m_2$&\multicolumn{2}{c}{with absolute orbit and $\pi$} && $f(m)$ &$m_i\sin^3 i$	& N	\\
\noalign{\smallskip}
\noalign{\smallskip}
$R_1, R_2$ & N &R$_i$\arcsec&&	\multicolumn{2}{c}{inferred from Sp. and L} &	$r_i=R_i/a$\\
\noalign{\smallskip}
$L_2/L_1$ & Y	& Y	&&	N & from Sp.& Y\\
\noalign{\smallskip}
limb dark. 	& N	& y  && N & N	& y \\
\noalign{\smallskip}
gravity dark.	& N	& y	&& N	& N	& y \\
\noalign{\smallskip}
\noalign{\smallskip}
\tableline
\end{tabular}
}
\end{center}
\end{table*}

Binaries are the primary source of fundamental stellar parameters and an important  
benchmark of theoretical models. Double-lined
eclipsing systems offer a {\em purely geometrical} means of determining stellar radii
and masses. 
 Besides, well before the amazing achievements of asteroseismology, binaries provided a
probe to stellar interiors through the study of apsidal motions and of
secular orbital evolution and synchronization. Basic information about the
stellar structure (mean densities, size of convective cores) can indeed
be derived from the long term study of their  best known parameter, the orbital period,
or from studies of period distribution and synchronization of suitable samples.

In recent years, double lined eclipsing binaries have  proven as well to be 
accurate distance indicators to clusters in our galaxy or even to the galaxies of the
local group, providing independent determinations and tests of the first rungs of the distance
ladder.
Moreover, the quick development of interferometric techniques is  opening the
possibility of  directly deriving  effective temperatures and atmospheric
properties of an increasing number of binary components, a very important
result, that will allow to ``close the loop" of fundamental parameter
determination. 

Finally it has to be stressed that, anyway, binaries cannot be  ignored, for the simple reason
that {\em multiplicity is the rule}, not the exception. Fifty to seventy percent of all stars,
at least in the solar neighborhood, are members of binary or multiple systems.
Therefore,  the formation and evolution of stars (and of clusters
and galaxies they populate) cannot be really understood without a deep knowledge of binary stars.

\section{Binaries to determine the fundamental stellar parameters}
  
We will distinguish  between the determination of stellar masses and radii, 
relying on pure geometrical methods, and the less direct one of effective temperatures. 

\subsection{Stellar masses and radii}
Conceptually, the simplest derivation of stellar masses stems from astrometric binaries
with known {\em absolute} orbits of the components (with respect to nearby stars) and known
parallax. 
This is, indeed, the only case in which the  mass determination can be achieved  by only one technique
(see Table 1). The table -- an updated version of the similar one from the
classical book of \citet{batten73} -- gives a summary of the parameters that can be derived
(or inferred) from astrometric binaries, (AB) subdivided into  visual (VB) and interferometric (IB) binaries,
double and single-lined spectroscopic 
binaries (SB2, SB1) and eclipsing systems (EB). The first seven parameters in the table define the orbit: 
semiaxis, $a$, eccentricity, $e$,  orbital period, $P$, epoch of primary 
minimum, $T_0$, inclination, $i$, longitude of periastron, $\omega$, and orientation of
the line of nodes, $\Omega$. The remaining parameters  are the component masses, radii,
ratio of luminosities (usually in a color band), and the parameters defining second order effects
in the light curve (limb and gravity darkening). The small `y' means `yes, in particular cases'
(such as, for instance, with high accuracy  observations). The value of $\Omega$ is ambiguous
by 180\deg\ if determined by visual observations alone.

IBs, similarly to VBs, provide a way to reconstruct 
the orbit and have as well a number of other advantages:
a larger limiting distance, the possibility of determining as well the (angular) stellar
radii and, in favorable cases, even the limb and gravity darkening 
 \citep[see, for instance, the comprehensive review of][]{quir01}.
\begin{figure}[ht]
\centerline{\includegraphics[width=8cm]{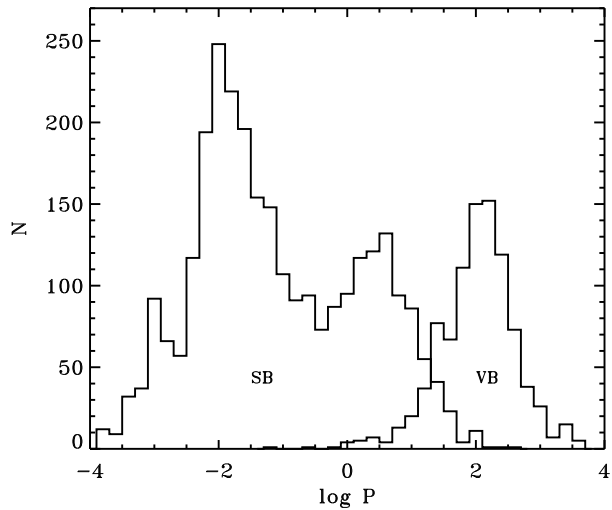}}
\caption[]{Period distribution from the SB$^9$ catalog of spectroscopic binaries
\citep{pourb05} and the 4$^{th}$ Catalog of Visual Binary Orbit \citep{wor83}}.
\label{sb9vb4}
\end{figure}
 Most of the main parameters can be determined from ABs of known absolute orbits, if the  the 
linear scale  is introduced by the parallax, $\pi$.  The system total mass 
directly descends, then, from the 3${^\mathrm{rd}}$ Kepler law, and the mass ratio from that of the orbital
semi-axes.  Absolute orbits, however, are available for very few systems.

Alternatively, a second relation involving  masses can be obtained from spectroscopy and, if 
the radial velocity amplitudes of both components are known (SB2), the linear scale is provided by the
direct determination of $a \sin i$,  so that  parallax is no longer needed. 
This applies to  a larger number of systems and  to larger distance (limited by astrometry), but the 
intersection of the two samples,
 VB $\cap$ SB, is still marginal, see Fig.\ref{sb9vb4}.
That is easily explained by the fact that ``typical" visual binaries (as those collected in catalogs)
were observed either by eye or by filar micrometers, and  the minimum measurable separation was around 0.2\arcsec. 
The  radial velocity amplitude of a nearby system with that separation, total  mass of 2 M$_\odot$,  and at a
distance of -- say --  10~pc  is in the range  $5\div 15$ Km~s$^{\mathrm{-1}}$, depending on the mass ratio 
( and  with circular orbit, i=90\deg and $q=0.2\div 1.0$). The quantity scales, for a given
angular separation, as $d^{-\onehalf}$.  

 Therefore, even if  pointed observations of specific targets with modern instrumentation
(such as optical interferometers and high resolution spectrographs on large telescopes)   
can provide minimum angular separation of a few milli-arcseconds (see below) and  radial
velocities $<1$ Km~s$^{\mathrm{-1}}$, there is no surprise for the small intersection 
of the two distributions of Fig.\ref{sb9vb4}.

Furthermore, the accuracy of the masses obtained by this means is typically not very high,
the main reason being  the dependency of the mass on
$\sin^3 i$, so that face-on (low inclination) system  do not allow precise determinations.
According to a study by \citet{pourb00} on eighty  ABs,  only 10\% of systems has masses
known to better than 3\%.  

The situation is, however, quickly evolving with the development of new generation interferometers.
To cite just one, striking, result: recently  \citet{bod05} have determined the masses of
12~Boo components with $\simeq 0.4 $ \% accuracy using the Palomar Testbed Interferometer (PTI) and
echelle spectroscopy. This  system has a period  $P=9.6^{\mathrm{d}}$and an apparent separation 
of only 3.45 mas (and $\pi=27.74$ mas).   

The latest development of long baseline interferometry, with new generation instruments already in
operation (such as the CHARA array)  or planned for next future, is pushing the resolution limit
to values well below the mas. The basic formula expressing the resolution of an interferometer,
$R=\lambda/2b$, where $b$ is the baseline and $\lambda$ the wavelength, tells us that to resolve, for instance, 
a spectroscopic binary with a separation of 40 R$_\odot$ and at a distance of 500 pc (i.e with an angular separation
of 0.4 mas) a baseline of $\simeq$ 300 m is needed, a value that is comparable, for instance,  to
the longest one of CHARA array.
 Space-born experiment (such as SIM, Stellar Imager) will, of course, go much beyond.

By now, however,  the ``royal road" to stellar mass determination is still traced by double-lined 
eclipsing binaries (eSB2).
The detailed analysis of detached eSB2s yields masses with an accuracy often better than 1\%, i.e.
suitable as  stringent tests for theory (according to \citet{and91} a 1\%-2\% accuracy in stellar 
parameters is necessary to really constrain the models).  
The main limitation, in this case, is  the rapid decrease of the eclipse probability 
with orbital period. For given masses and radii this quantity scales approximately as $P^{-4/3}$
\citep{mr99}.  

  According to the results of photometric surveys, as OGLE, MACHO, Vul\-can, STARE,  ASAS, 
 0.1--0.2 \% of stars are  eclipsing binaries:  our galaxy 
should contain, therefore, $\sim 10^8$ EBs. Only two hundred systems, however, have been studied in detail
(out of $\sim 10000$ presently known), and for only half of those accurate parameters could be
determined \citep{and91,and98,and02}. Some stellar types  are dramatically under-represented among EBs
(low and high mass stars, giants), though the situation is steadily improving, thanks to the
results of the abovementioned surveys.

Eclipsing spectroscopic binaries provide as well precise (at a few percent level) stellar radii. The great 
advantage is that the radii (as the masses) determined by eSBs are distance independent.
 An alternative source is, also in this case, long baseline optical interferometry. This requires, anyway, the transformation 
to the linear scale of the measured angular diameters. More than hundred stellar radii determinations can be 
found in the literature: obtained by the Narrabri Intensity Interferometer and VLTI \citep{hanb74,dif04} for hot stars,  by the Mark III,
 NPOI (Naval Prototype Optical Interferometer) and PTI \citep{moz03,nor99,lan01} for cooler stars. 
If the object under study is a spectroscopic binary (or has a known parallax) the linear diameters can be obtained.
The accuracy  of angular diameters by interferometric measurements can vary from 0.5\% to a few or
several percent, depending on the system characteristics and the wavelength (redder wavelengths provide more accurate
results). In general the binary case is more difficult to treat than that of single stars.   

To be noticed that interferometric and eclipsing binaries results are somewhat complementary, the sample of 
stars studied  by interferometry contains, in fact, many giants which are scarce among EBs. 

\subsection{Stellar temperature, luminosity and distance}
	For comparison with stellar models or distance estimate the effective temperatures are needed.
These can be obtained by different methods:
\begin{itemize}
\item  empirical calibrations and absorption free color indexes (e.g. from Str\"{o}mgren or IR photometry).
The distance is derived from the distance modulus by the standard expression:
\begin{equation}
(m_{\mathrm{V}}-M_{\mathrm{V}})_0=m_{\mathrm{V}}-A_{\mathrm{V}}-M_{bol,\odot}+ 5 \log \frac{R}{R_\odot}+
10 \log \frac{T_{\mathrm{e}}}{T_{\mathrm{e},\odot}}+\mathrm{BC}.
\label{distmod}
\end{equation}
The typical uncertainties of the various quantity appearing in Eq. \ref{distmod} were  estimated by \citet{clau04},
in the application to distance estimation of the LMC,
and imply a distance modulus accuracy  of 0.10-0.15 magnitudes, with somewhat better results for early
type stars (0.07 -- 0.09 mag). These are characterized by a favorable behavior of the bolometric correction with 
temperature: a linear fit of the early star region of the \citet{flo96} ($\log T_{\mathrm{e}}$--BC) calibration yields 
BC$\propto -5.4 \log T_{\mathrm{e}}$, so that the  dependence of their distance modulus on $T_{\mathrm{e}}$
is weaker than appearing from Eq.\ref{distmod}.
\item for SB2-EBs: fit of the UV -- optical SED,  avoiding the use (and uncertainties) of calibrations \citep{gui98, fitz99}.  
The measured flux, $f_{\lambda,\oplus}$, can be expressed as function of the stellar radii, the surface fluxes,
$F_{\lambda,i}$, the distance, the absorption coefficient and color excess:  
\begin{equation}
f_{\lambda,\oplus}=\left( \frac{R_1}{d} \right )^2 \left[ F_{\lambda,1}+ \left(\frac{R_2}{R_1} \right)^2 F_{\lambda,2} \right ] 
\cdot 10^{-0.4[E(\lambda-V)+A_{\mathrm{V}}]}
\label{fitsed} 
\end{equation}
The surface fluxes, $F_{\lambda,i}=f(T_e, \log g, m/H,\mu)$ can be derived from model atmospheres (the surface gravities are
known, the microturbulence velocity, $\mu$, is held fixed). A best fit for $T_e$, $ m/H$ (metallicity), $ A_\mathrm{V}$,
$ E(\lambda-V)$, $d$ provides the
temperatures, the  characteristics of interstellar absorption and the distance. 
The method has been applied with  success to several extragalactic eSB2s. The procedure is rather complex,
requires multi-wavelength data and is better applied to hot stars  in a temperature range 
free from NLTE effects, which can be a problem for model atmospheres  (i.e below 30000 K).  The results
can be, however, of great accuracy.   For instance, the effective temperature of the HV~2274 primary component, as obtained 
by \citet{gui98}, is $T_e=23000 \pm 180$ K.
\item The method of infrared fluxes \citep{bla77,bla94}  is based on the weak dependence of the IR flux on 
$T_{\mathrm{e}}$ and benefits of the negligible interstellar absorption at IR wavelengths. In the single star case,
an iterative process starting from  a  guessed $T_{\mathrm{e}}$  and model atmospheres (to compute the surface
flux $F_{IR}$) provides, because of the abovementioned weak dependence,  a precise value of the ratio: 
\begin{equation}
\frac{f_{IR,\oplus}}{F_{IR}}=\frac{\theta}{4}^2
\label{irfl}
\end{equation}
 That  yields, in its turn,  the value of the angular diameter, 
$\theta$. An improved value of $T_{\mathrm{e}}$ can then be obtained from the measured integrated 
flux at Earth, by means of the relation:
\begin{equation}
f_{\mathrm{bol,\oplus}}=\frac{\theta^2}{4} \sigma T_{\mathrm{e}}^4
\label{bolf}
\end{equation}
from which a new iteration can be started.
A straightforward extension of this method to binary stars has been formulated by \citet{sma93}. 

The method yields late type star temperatures determined with 1.3\% accuracy \citep{ram05} and, as well, angular
radii that agree within 1\% with interferometric determinations \citep{nor01}.
It is obvious from Eq. \ref{bolf} that, in the case of nearby stars, the effective temperatures can be directly derived
from the interferometric angular diameters  and bolometric fluxes. 
\item The detailed spectral analysis after applying, if necessary, an algorithm for component spectra
disentangling \citep{had95}.  
\end{itemize}

\section{Using the binary tool}
For reasons of space it is impossible to examine in detail all the ways 
binaries can be used as effective ``astrophysical tools".
I will focus, therefore, on their effectiveness as benchmark for stellar models
and on their recently acquired  role of independent contributors to the first rungs of the distance scale.
Other important topics, such as the study of apsidal motion, of spin--orbit synchronization
and circularization, yielding precious information on  stellar structure, have been
reviewed by several authors in a recent conference \citep{cla05}.
\subsection{Binaries as theoretical model benchmarks}
\begin{figure}[ht]
\centerline{\includegraphics[width=8cm]{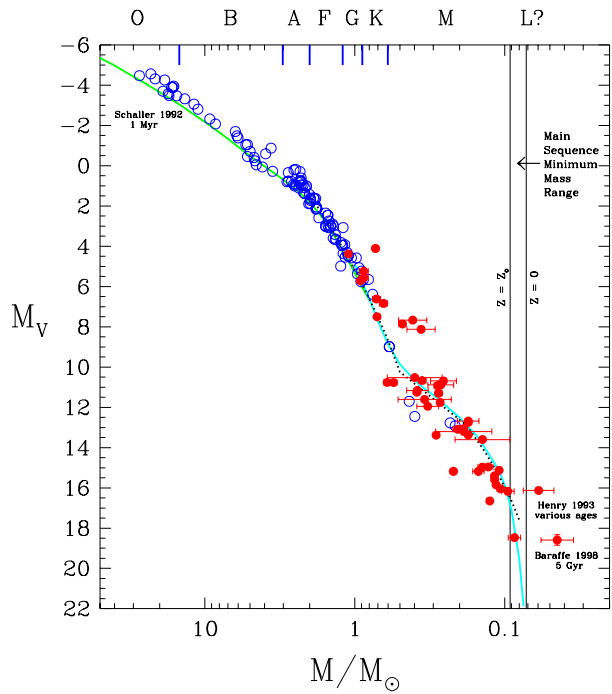}}
\caption[]{The Mass-Luminosity relation for field stars according to Henry (2004). Open points
represent eclipsing binaries, solid points astrometric binaries. The fit
in the massive star region  is from \citet{scha92}, in the low mass one from \citet{baraffe98}. 
An empirical fit (dotted line)  from \citet{henmc93} and \citet{henry99} is also shown.}
\label{ml}
\end{figure}

Fig \ref{ml} shows the mass -- luminosity relation of \citet{henry04} 
derived from binaries with accurate elements.  The upper part of the plot is populated by
EBs, the lower one essentially by ABs (with larger errors). Only a handful of
low mass eSB2s are known. At the other end, the most massive stars have 
$M\cong 30$ M$_\odot$. It has to be mentioned that the mass record actually belongs to 
the recently analysed  double-lined eclipsing binary WR20a, member of the open cluster Westerlund 2. 
The system is formed by Wolf-Rayet components of $83 \pm 5 $ and $ 82 \pm 5$ M$_\odot$ \citep{bon04},
the largest directly determined masses.
AS the figure collects only MS binaries,  neither WR20a nor the few known eSB2 with 
(sub)giant components are included. 
 
Fig. \ref{ml} shows a fair  agreement between theoretical models and observations,
but the data cannot be used to really constrain the models, because of the intrinsic scatter
due to age and metallicity effects. Besides, the most stringent constraints are those
posed by  masses and radii, or surface gravities.  
When accurate values of these quantities are compared with  standard theoretical models 
several discrepancies appear.

For instance, \citet{and98} showed that the surface gravities (derived with precision better
than 2 \%) of a sample of  intermediate mass MS eSB2 components   are smaller
than the values expected from standard models (see Fig.1 of the abovementioned paper).
That is explained in terms of the inadequacy of the treatment of internal mixing:
to fit the observations a  convective core larger than that of standard models is needed, 
implying a certain amount of extra mixing. By the way, models with larger convective core are 
more centrally condensed, in agreement, as well, with the results from the studies 
of apsidal motions.
This requirement has been modeled, in absence of
a physically consistent description, by  introducing core overshooting in a parametric form. 
 The obvious disadvantage of the parametric treatment is that the physical model
cannot be tested, though in some cases very high values of the overshooting parameter
are obtained which  are difficult to accept.
 
\begin{figure}[ht]
\centerline{\includegraphics[width=10cm]{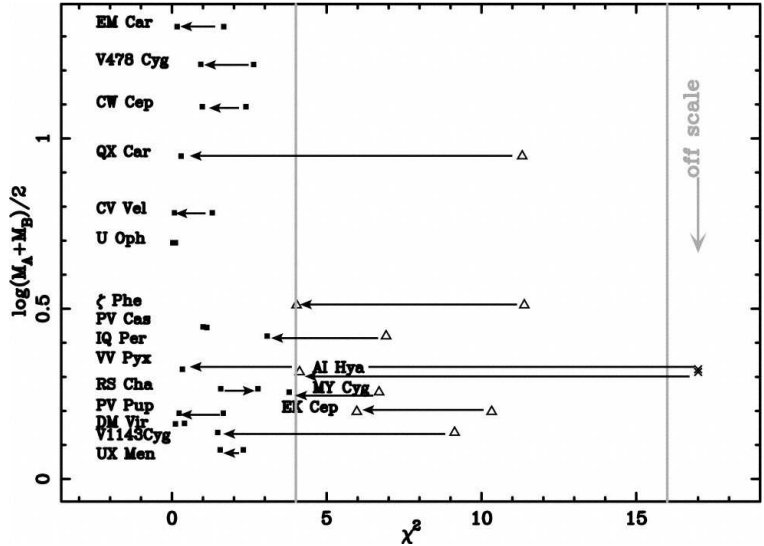}}
\caption[]{The $\chi^2$ values for best models of binaries vs the mean
system mass from \citet[][courtesy of ApJ]{you05}. Arrows are drawn from the $\chi^2$ values obtained
by \citet{you01} with standard models by the same code,  to the values 
from the updated 2005 version, including extra-mixing by internal waves. 
The vertical line at $\chi^2=4$ indicates a fit in which the models fall
inside -- in the $\log R - \log L$ plane -- the observational error boxes of both components.}
\label{chisq}
\end{figure}

A different approach has recently been proposed by \citet{you05}, who  find a very
good agreement between the observed parameters (radii and luminosities) of a subset of 
the \citet{and91} sample, made out of 18 eSB2s with MS components and masses in the range 
$ 1.2 \leq M \leq 30$ M$_\odot$, and the models from their evolutionary code TYCHO.  
The last version of TYCHO includes a consistent
treatment of mixing by internal waves, which   has on the stellar structure an effect 
similar  to overshooting, but introduces no free parameter.
The fit, performed by a $\chi^2$ algorithm (in the hypothesis of equal age for
both components) greatly improves when  extra mixing is taken into account,
all stars fall within the error boxes in the $\log R - \log L$ diagram, see
Fig.\ref{chisq}. 

A second, still open, issue concerns the results of testing low mass models. Only in the
last years a (still small) number of eSB2s has been added to the two well known 
calibrators of the lower MS, YY Eri and CM Dra, often thanks to the completeness
of variability searches of photometric surveys. These are:
CU Cnc \citep{rib03}, GU Boo \citep{rib05}, OGLE BW3V38 \citep{mm04} and 
TrES-Her0-07621 \citep{cre05}. 
  The study of these systems has revealed the existence of a systematic
discrepancy between models and observations, in the sense that real 
late K--M dwarfs seem to have larger radii (up to 20 \%)
and lower effective temperatures (by as much as 150 K) than current models
(see Ribas contribution, these proceedings). Among the suggested
explanations the most likely is the presence of strong magnetic fields. 
The non-standard stellar models of \citet{MMD01}, which include a simplified treatment of magnetic
field effects, suggest that late-type active stars  (hosting magnetic fields 
up to several tens of MGauss at the base of the convective zone)
should have, indeed, larger radii and lower temperatures than similar, inactive,
dwarfs.
In this case as well binaries point out the need of  non-standard 
ingredients in  stellar models.

\subsection{Photometric surveys and rare binaries}
The photometric surveys for variability search as OGLE, MACHO  and their related programs
\footnote{ a quite complete documentation is available at http://wwwmacho.mcmaster.ca/ 
and http://sirius.astrouw.edu.pl/$\sim$ogle/}
yielded, indeed, many by-product discoveries of interesting and rare binary systems,
including a few eclipsing binaries with pulsating components   
(that could provide an independent determination of the pulsating component mass 
with spectroscopy follow-up). 
Three candidate eclipsing binaries
containing an RR~Lyr component were discovered by OGLE-III  in the LMC \citep{sos03}, while three 
EBs with Cepheid components in the MACHO database of LMC variable stars have been studied  
in detail by \citet{alc02}.
\begin{figure}[ht]
\centerline{\includegraphics[width=8cm]{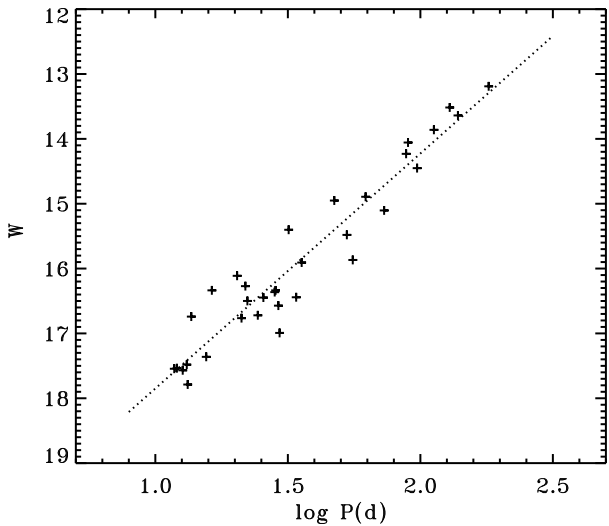}}
\caption[]{The relation between the Wesenheit index, W, and orbital period for the
35 binaries with ellipsoidal giant primary extracted by \citet{rm01} from the OGLE-II
SMC catalog of variable stars.}
\label{ellg}
\end{figure}

Besides, a new class of interesting binaries, with periods in the range $10\div 180^{\mathrm{d}}$, and whose light 
curves are produced just by the ellipsoidal variation of a giant primary, was identified by \citet{rm01}
in the SMC fields observed by the OGLE-II experiment. These objects, thanks to the scaling with period of
an essentially fixed geometrical
configuration (primary in contact with the Roche lobe), follow a period/color/luminosity relation and
can be used as auxiliary (but independent) distance indicators. Fig.\ref{ellg} expresses this
property trough the relation between the orbital period and the Wesenheit absorption-free color index, 
$\mathrm{W=I-1.55(V-I)}$, generally used for pulsating stars. In terms of absolute V magnitude and
color index the P-L-C relation can be written as:
\begin{equation}
M_V =  -3.43 \, \log P + 2.04 \, (V-I)_0 + 2.80, \; \; \sigma=0.34 
\label{elleq}
\end{equation}
where $\sigma$ is the standard deviation of the fit of the 33 systems with ellipsoidal giants 
of \citet{rm01}. 
The same type of eclipsing binaries were, 
later on,  detected in large number in the LMC  \citep{sos04}. The brightest of the LMC samples obey 
a  relation similar to that in Fig. \ref{ellg}, differing for a shift by an amount close to the 
difference in distance modulus between the  Clouds.
 
\subsection{Binaries and the distance ladder}
 As shown in the previous sections, the detailed analysis of detached eSB2s is in itself a powerful tool 
to measure distance. The idea of using binaries as distance indicators to clusters and to the
closest members of the Local Group dates back, indeed, to the first half of the 20th century
\citep{gap40}, i.e. about  
thirty years after the discovery by \citet{leav1908} of eclipsing binaries in the Magellanic Clouds.
 It was necessary, however, to wait some decades more to achieve the necessary technical and instrumental
development. The first modern and accurate determination of the LMC distance modulus  by a
detached eSB2, the Harvard Variable HV2274, is due to \citet{gui98}, previous attempts  having
essentially an historical interest. 

Since then, a few other systems have been studied, providing clues of a non-negligible depth of the 
Cloud along the line of sight \citep{gui04}. 
A parallel analysis of binaries  in the Small Magellanic Cloud has been undertaken by
\citet{hhh03}.
 
 In both cases the binary-based results  are in excellent agreement with the determinations by other
means and have an accuracy comparable to that of the best indicators. The latest estimate for the
LMC, according to \citet{gui05} is $(m-M)_0=18.42 \pm 0.07$, in good agreement with the value to which the
various other results seem to converge (the weighted average from different methods,
according to \citet{alv04} is $18.50\pm0.02$).

 Similarly, the SMC distance modulus, as obtained by \citet{hhh03,hhh05} from fifty EBs, is 
 $(m-M)_0=18.91 \pm 0.03 \pm 0.1$
(the last term  being an evaluation of systematic errors). The value is in between the `` long" distance
derived from Cepheids \citep[see, e.g., ][]{bono01} and the ``short" one from red clump stars 
\citep[e.g][]{twa99}.   

With the availability of 10m-class telescopes, it is now feasible to get the first spectroscopic
observations of eSB2s in galaxies of the Local Group.  Several hundreds  EBs have been detected
by photometric surveys in M31, a few tens in M33 \citep{bon03,mac04}. A few EBs have been found as well in the
smaller members of the local group: Fornax, Leo, Carina, NGC~6822 
\citep[see][and references therein] {gui04}, and Phoenix \citep{gal04}. The first, preliminary,
 determination of the distance to M33
by means of an eSB2 has been recently obtained \citep{bon05}. 
No doubt therefore that this
field, in very rapid development, will yield fundamental contributions to the 
knowledge of  nearby galaxies and of the Universe.

\section{Conclusions}
The fact that in our galaxy the majority of stars are components of binary or
multiple systems is for us a great stroke of luck, as we got efficient
and versatile tools to increase our knowledge of stars in the Milky Way and other 
galaxies.
 
 The technical developments of the last decades has opened the possibility of
studying binaries in the Magellanic Clouds, in M31 and in M33, and of accurate 
determinations of  their absolute elements and distance. On one side, this allows to
study different stellar populations from those of our galaxy, on the other
it provides an independent and scale free distance indicator. The results from 
binaries for the distance to the LMC, which is responsible of a non negligible contribution to
the error budget in the determination of the Hubble constant, are of great value,
 being free from zero-point uncertainties. 
While at present accurate distance determinations  are possible only
for the Magellanic Clouds, because of the magnitude limits for spectroscopy,
in the next future the other members of the
Local Group, in which many eclipsing binaries have already been detected,
will become reachable. The first example in this direction is the first determination
of the distance to M33 obtained by spectroscopy with the  Keck telescope. 

   The large databases and catalogs from microlensing and planet-search surveys
of the last decade have already provided light curves of about 15000 eclipsing
binaries.
Many  interesting and rare system were found  and a complete analysis 
was possible after follow-up observations.
The next future reserves an increase of two more order of magnitudes, millions
of light curves will be provided by wide-field all-sky surveys, and at 
different wavelengths. It will be imperative, therefore, to develop ad-hoc
automatic algorithms for data reduction and analysis.
  
  Space experiments devoted to asteroseismology and planet searches (such as
Corot and Kepler) will provide, as well, extremely precise and  continuous photometry (at 
a level of 10$^{-4}$ mag and  on baseline of months or years) of many thousand  binaries.
Detailed study of the stellar atmospheres (from, e.g. the direct determination of
the limb darkening from the light curves) will be possible,  together with a
long term monitoring of stellar activity phenomena. Certainly many new objects 
belonging to the present groups of rare binary systems will be found. 
 
    Finally, interferometry is  already filling the gap between spectroscopic 
and visual binaries, providing more and more accurate stellar parameters
for closer and closer systems. In the next future close binary systems will
be routinely resolved by long baseline optical interferometers and space
missions, as Stellar Imager, will be able to resolve surface features on
the individual components.

 Therefore, the classical field  of research on binary stars has  still an extremely 
bright future, promising  great developments and interesting research topics
for  you all.  

\acknowledgements
I thank the organizers,  Conny Aerts and Tibor Hegedus, for 
their invitation to this very special PhD conference, where I could
meet and interact with many enthusiastic young people working on 
variable stars. 


\end{document}